\documentclass[a4paper,11pt]{article}
\pdfoutput=1
\usepackage{jheppub} 
\usepackage{graphicx,color,rotating}
\usepackage{hyperref}
\usepackage{epsfig,color}
\usepackage{slashed}
\usepackage{amsfonts}
\usepackage{ulem}
\usepackage{color}
\newcommand{\lsim}{\raisebox{-0.13cm}{~\shortstack{$<$ \\[-0.07cm] $\sim$}}~}
\newcommand{\gsim}{\raisebox{-0.13cm}{~\shortstack{$>$ \\[-0.07cm] $\sim$}}~}
\graphicspath{{D:/Desktop/draft}}
\usepackage{tabularx}
\usepackage{graphicx}
\usepackage{adjustbox}
\usepackage{tabularx}
\def\lsim{\:\raisebox{-0.5ex}{$\stackrel{\textstyle<}{\sim}$}\:}
\def\gsim{\:\raisebox{-0.5ex}{$\stackrel{\textstyle>}{\sim}$}\:}
\usepackage{float}
\usepackage{subfig}
\usepackage{amsmath,mathtools}
\begin{document}
	\author{
	Po-Yan Tseng$^{1,2}$ and Yu-Min Yeh$^{1}$}
	\affiliation{
		$^1$ Department of Physics, National Tsing Hua University,101 Kuang-Fu Rd., Hsinchu 300044, Taiwan, R.O.C. 
	}
	\affiliation{$^2$ Physics Division, National Center for Theoretical Sciences, Taipei 106319, Taiwan R.O.C.}
	\date{\today}

\abstract{
We investigate a neutrino-scalar dark matter (DM) 
$\nu\phi$ interaction encountering distinctive neutrino sources, namely Diffuse Supernova Neutrino Background (DSNB) and Active Galactic Nuclei (AGN).
The interaction is mediated by a fermionic particle $F$, in which the $\nu\phi$ scattering cross section characterizes different energy dependent with respect to the kinematic regions, and manifests itself through the attenuation 
of neutrino fluxes from these 
sources.
We model the unscattered neutrino flux from DSNB
via core-collapse supernova (CCSN) and star-formation rate (SFR), 
then incorporate the present Super-Kamionkande and future DUNE/Hyper-Kamiokande experiments to set limits on DM-neutrino interaction.
For AGNs, NGC 1068 and TXS 0506+056, where the neutrino carries energy above TeV, 
we select the kinematic region $m^2_F \gg E_\nu m_\phi \gg m^2_\phi$ such that the $\nu \phi$ scattering cross section features an enhancement at high energy.
Furthermore, taking into account the DM spike profile at the center of AGN, we constrain on $m_\phi$ and scattering cross section 
via computing the neutrino flux at IceCube, where the $\phi\phi^*$ annihilation cross section is implemented to determine the saturation density of the spikes. Notice that the later results heavily rely on the existence of DM spike at the center of AGN, otherwise, our results may alter.}

\title{Phenomenology of Neutrino-Dark Matter Interaction in DSNB and AGN}

\maketitle

\section{Introduction}
Dark Matter (DM), accounts for $27\%$ of the Universe, is yet identified from particle physics point of view. 
DM particles have not been detected in the lab if they have only revealed the interaction with Standard Model (SM) particles through gravity which provided indirect evidence of DM in cosmological scale. 
Many observations have been proposed to study the imprints of fundamental interaction between DM and Standard Model (SM) particles, such as the cosmological and astrophysical effects on various hypothetical DM interactions~\cite{Mangano:2006mp,Wang:2023csv,Hooper:2021rjc,McMullen:2021ikf,Farzan:2014gza,Akita:2023yga,Ferrer:2022kei,Fujiwara:2023lsv}, 
or the boosted DM scenarios where non-relativistic DM particles gain energy through the upper scattering with cosmic electrons or neutrinos~\cite{Ghosh:2021vkt,Zhang:2020nis,Jho:2021rmn,Agashe:2014yua,Das:2024ghw,Zapata:2025huq}.
%

In this work, we focus on the hypothetical interaction between scalar DM $\phi$ and neutrino then study the phenomenological outcome.
%
%
In particular, we introduce fermion-mediated interactions in Section \ref{nuphiinteraction}, where the $\nu \phi$ scattering cross section $\sigma_{\nu \phi}$ follows different neutrino energy dependence according to various kinematic regions.
%
For example, in the limit $E_\nu \gg m_\phi \simeq m_F$, $\sigma_{\nu \phi}$ is inversely proportional to $E_\nu$, and thus the low neutrino sources are more applicable.
On the other hand, in the heavy mediator limit, $m^2_F \gg E_\nu m_\phi \gg m^2_\phi$, the $\sigma_{\nu \phi} \propto E_\nu$ exhibits enhancement at high energy. 
%
Therefore, we consider two neutrinos sources: The Diffuse Supernova Neutrino Background (DSNB) and the Active Galactic Nuclei (AGN). These two special sources provide prolific neutrinos with energy of $\mathcal{O}(10\,{\rm MeV})$ and $\mathcal{O}(100\,{\rm TeV})$, respectively. 
The existence of neutrino-DM interactions would manifest from attenuating the neutrino flux during the propagation from the sources to the Earth.

The anticipated DSNB originating from the distant Core-Collapse SuperNova (CCSN) is not yet confirmed discovery by the current neutrino detectors. It is pointed out that it can be potentially observed at Super-Kamiokande/Hyper-Kamiokande (SK/HK) 
\cite{Beacom:2010kk,Horiuchi:2008jz,DeGouvea:2020ang,Super-Kamiokande:2013ufi}.
In Section \ref{DSNB}, we model the DSNB flux of electron neutrino via a thermal non-degenerate Fermi-Dirac distribution with temperature 
$6.6\,{\rm MeV}$ \cite{DeGouvea:2020ang}
and focus on the open energy region of $\mathcal{O}(10\, {\rm MeV})$  sandwiched by the overwhelming backgrounds expected at the future neutrino detectors (i.e reactor $\bar{\nu}_e$ from beta decay and $\nu_e$ from inverse muon decay). 
The DSNB flux depends on the rate of CCSN, which relates to the history of Star-Formation Rate (SFR).  
We compute the DSNB flux via including the effect of $\nu\phi$ scattering, 
and perform the sensitivity analysis by introducing the present SK~\cite{Super-Kamiokande:2021jaq} and future HK/DUNE (Deep Underground Neutrino Experiment)~\cite{Hyper-Kamiokande:2022smq}.
The DUNE is the future neutrino detection project and aims to investigate various topics in neutrino physics, such as neutrino oscillations, baryon number violation, supernova neutrino bursts, etc.~\cite{DUNE:2020lwj,DUNE:2020ypp}
To obtain corresponding event number from DSNB, we calculate the MeV neutrino-Argon, $\nu_e{\rm Ar}$, scattering rate and estimate the sensitivity on $\nu\text{-}\phi$ coupling $y$.

AGNs are considered as alternative sources emit neutrinos above $\mathcal{O}$(TeV), which is ideal to probe the kinematic region where the neutrino-DM cross section increases with energy.
The IceCube Collaboration has observed the ultra high energy neutrinos from the galaxies, NGC 1068 and TXS 0506+056 \cite{IceCube:2022der,IceCube:2018cha,IceCube:2018dnn}.
It was assumed in earlier works~\cite{Gondolo:1999ef,Herrera:2023nww,Cline:2023tkp}
that AGN possess a spike-like DM density profile around the center SuperMassive Black Holes (SMBH). 
Furthermore, this profile, especially the saturation DM density, is directly associated with the DM self annihilation.
It is legitimate to assume emitted neutrinos interact intensively with the DM around the SMBH, which increases the neutrino attenuation.
Considering the interplay between the $\nu\phi$ scattering and the $\phi\phi^*$ annihilation cross sections on NGC 1068 and TXS 0506+056,
we demonstrate the upper bounds on $\nu\phi$ interaction in Section~\ref{agn}.


We conclude our results in Section~\ref{conclusion} and show the detailed calculations for scattering cross sections and spike-like density parameters in appendices.

\bigskip

\section{Scalar Dark Matter-Neutrino Interactions}
\label{nuphiinteraction}
We consider neutrino as Majorana fermion which couples to scalar DM $\phi$ through a fermionic mediator $F$~\cite{Boehm:2003hm}:
\begin{equation}\label{newinteraction}
		\mathcal{L}_{F{\rm -med}}=y(\phi\,\overline{\nu_L} F_{R}+\phi^\ast\,\overline{F_R}\nu_L)\,,
	\end{equation}
where $y$ is the coupling constant. 
The above effective interactions can be realized under the UV complete Majoron model which includes additional $SU(2)_L$ singlets, right-handed neutrinos, and Majoron (complex scalar)~\cite{Chikashige:1980ui,DiBari:2021dri}. The active neutrinos couple to right-handed neutrinos through the Higgs doublet field, meanwhile the Majoron only couples to the right-handed neutrinos. However, after the spontaneous symmetry breaking, certain among of mixing between CP-even components of Higgs doublet and Majoron would lead to the above effective interaction.
Here, we adopt the ``non-self conjugate'' DM ($\phi\neq\phi^\ast$) case,
because the "self-conjugate" dark scalar ($\phi = \phi^\ast$) does not contribute to the elastic scattering with neutrinos, but "non-self conjugate" case has $u$-channel contribution. The total cross section of $\nu\phi$ scattering is given in Appendix~\ref{app:A}, Eq.~(\ref{nuphicrosssection2}).
The neutrino-DM scattering cross section from Eq.~(\ref{newinteraction}) exhibits different energy dependent. For instance, when the neutrino energy $E_\nu$ is much smaller than the DM mass $m_\phi$ and the mediator mass $m_F$, we have
\begin{equation}\label{limit1}
\sigma_{\nu\phi}\simeq \frac{y^4E_\nu^2}{16\pi(m^2_F-m^2_\phi)^2}
\end{equation}
in case of $m_F>m_\phi$. For $m_F=m_\phi\gg E_\nu$, the cross section becomes
\begin{equation}
\sigma_{\nu\phi}\simeq \frac{y^4}{64\pi m_\phi^2},
\end{equation}
which becomes energy independent.
In contrast, for $E_\nu\gg m_{\phi,F}$, it can be approximated as
\begin{equation}\label{limit3}
\sigma_{\nu\phi}=\frac{y^4}{64\pi}\bigg[\frac{\ln\left(1+\frac{2E_\nu m_\phi}{m^2_F}\right)-1}{E_\nu m_\phi}\bigg]\sim E^{-1}_\nu.
\end{equation}
Another useful limit is when $m_F^2\gg E_\nu m_\phi\gg m^2_\phi$, the cross section becomes linear in $E_\nu$: \begin{equation}\label{parametrized}
	\sigma_{\nu\phi}\simeq \sigma_0\frac{E_\nu}{E_0},~~~~{\rm where}~~\sigma_0\equiv\left(\frac{y^4m_\phi }{32\pi m^4_F}\right)E_0\,.
\end{equation}
and $E_0$ is a arbitrary rescale energy.

For DSNB, the neutrinos are emitted with energy scale of $\mathcal{O}({\rm MeV})$. In this case, we choose similar masses of $m_\phi$ and $m_F$, in particular fixing $m_F/ m_\phi=1.1$ to avoid divergence in Eq.~(\ref{limit1}), and examine the relations between $y$ and $m_\phi$.
On the other hand, considering $\mathcal{O}({\rm TeV})$ neutrino sources, for instance AGN, the $\nu\phi$ scattering cross section with linear energy dependent is more applicable.
The aforementioned interactions cause the deduction of the anticipated neutrino fluxes at detectors, say DUNE or IceCube, which we will further discuss in the following sections.

To quantify the flux attenuation from distance neutrino sources, 
we need to calculate the transmittance $T$, defined as the ratio of the received and the emitted flux, which can be obtained from the optical depth $\tau$~\cite{Doring:2023vmk}
\begin{equation}
\label{taured}
	T=e^{-\tau},~~~~{\rm where}~~\tau(E_\nu,z)=\int^z_0{\frac{\Gamma(E_\nu,z^\prime)}{(1+z^\prime)H(z^\prime)}dz^\prime}\,,
\end{equation}
and scattering rate $\Gamma=\sigma_{\nu\phi}n_{\rm DM}$.\footnote{Obtaining the exact neutrino flux attenuation, the cascade equation needs to be solved~\cite{Cline:2024udd}, which takes into account two effects: neutrinos be scattered out of beam direction, and high-energy neutrinos be downscattered to lower energy. 
In the numerical analysis, we adopt the package $\nu$FATE which solves the cascade equation of neutrino flux and calculates the flux attenuation~\cite{Vincent:2017svp}. Particularly, we implement our $\nu\phi$ cross section and differential cross section from Appendix~\ref{app:A} into $\nu$FATE code according to Eq.(6) of Ref.~\cite{Vincent:2017svp}.}
The extragalactic averaged DM density is $\rho_{\rm DM}=n_{\rm DM}m_\phi=1.27\,{\rm GeV}\,{\rm m}^{-3}$. Here, the upper and lower integral limits associate with the redshifts of neutrino source and observer, respectively. With the redshift at $z^\prime$ included, we must replace $E_\nu$, the incident neutrino energy at today, by $E_\nu(1+z^\prime)$ and $n_{\rm DM}$ by $n_{\rm DM}(1+z^\prime)^3$. The Hubble rate at matter-dominating epoch is given by
\begin{equation}\label{hubble}
	H(z)=H_0\sqrt{\Omega_m(1+z)^3+\Omega_\Lambda}
\end{equation} 
with $H_0=67.36\,{\rm km}\,{\rm s}^{-1}\,{\rm Mpc}^{-1}$ and $\Omega_m=0.3153$, $\Omega_\Lambda=0.6847$ are the matter and vacuum contributions to the energy density \cite{Das:2021lcr,Planck:2018vyg}. 

\bigskip

\section{Diffuse Supernova Neutrino Background and DUNE Experiment}\label{DSNB}	
The DSNB models isotropic neutrino and antineutrino sources from core-collapse supernovae.
The distribution of neutrino emitted from a supernova has the following Fermi-Dirac form \cite{Beacom:2010kk}:
	\begin{equation}
	F_\nu(E_\nu)=\frac{E^{\rm tot}_\nu}{6}\frac{120}{7\pi^4}\frac{E^2_\nu}{T^4_\nu}\frac{1}{\exp(E_\nu/T_\nu)+1},
	\end{equation}
where $E^{\rm tot}_\nu=3\times 10^{53}\,{\rm erg}$ 
is the total emitted energy ($1/6$ factor stands for energy of one of $\nu_e$, $\nu_{\bar{e}}$, $\nu_\mu$, $\nu_{\bar{\mu}}$, $\nu_\tau$, $\nu_{\bar{\tau}}$) and $T_\nu$ is the neutrino temperature.
The diffuse differential neutrino flux without DM attenuation is given by
	\begin{equation}\label{DSNBflux}
	\frac{d\Phi_\nu}{dE_\nu}=\int_{0}^{z_{\rm max}}{\frac{R_{\rm CCSN}(z)F_\nu(E_\nu,z)}{H(z)}dz}.
	\end{equation}
The rate $R_{\rm CCSN}$ and the SFR parameters are \cite{Horiuchi:2008jz,Yuksel:2008cu}
	\begin{subequations}\label{rccsn}
	\begin{align}
	R_{\rm CCSN}(z)&=\dot{\rho}_\ast(z)\frac{\int_{8M_\odot}^{50M_\odot}{\psi(M)dM}}{\int_{0.1M_\odot}^{100M_\odot}{M\psi(M)dM}},\label{rccsn}\\
	\dot{\rho}_\ast(z)&=\dot{\rho}_0\left[
	(1+z)^{-10\alpha}+\left(\frac{1+z}{B}\right)^{-10\beta}+\left(\frac{1+z}{C}\right)^{-10\gamma}
	\right]^{-1/10}\label{sfr},
	\end{align}
	\end{subequations}
 \begin{figure}
    \centering
    \includegraphics[width=0.75\linewidth]{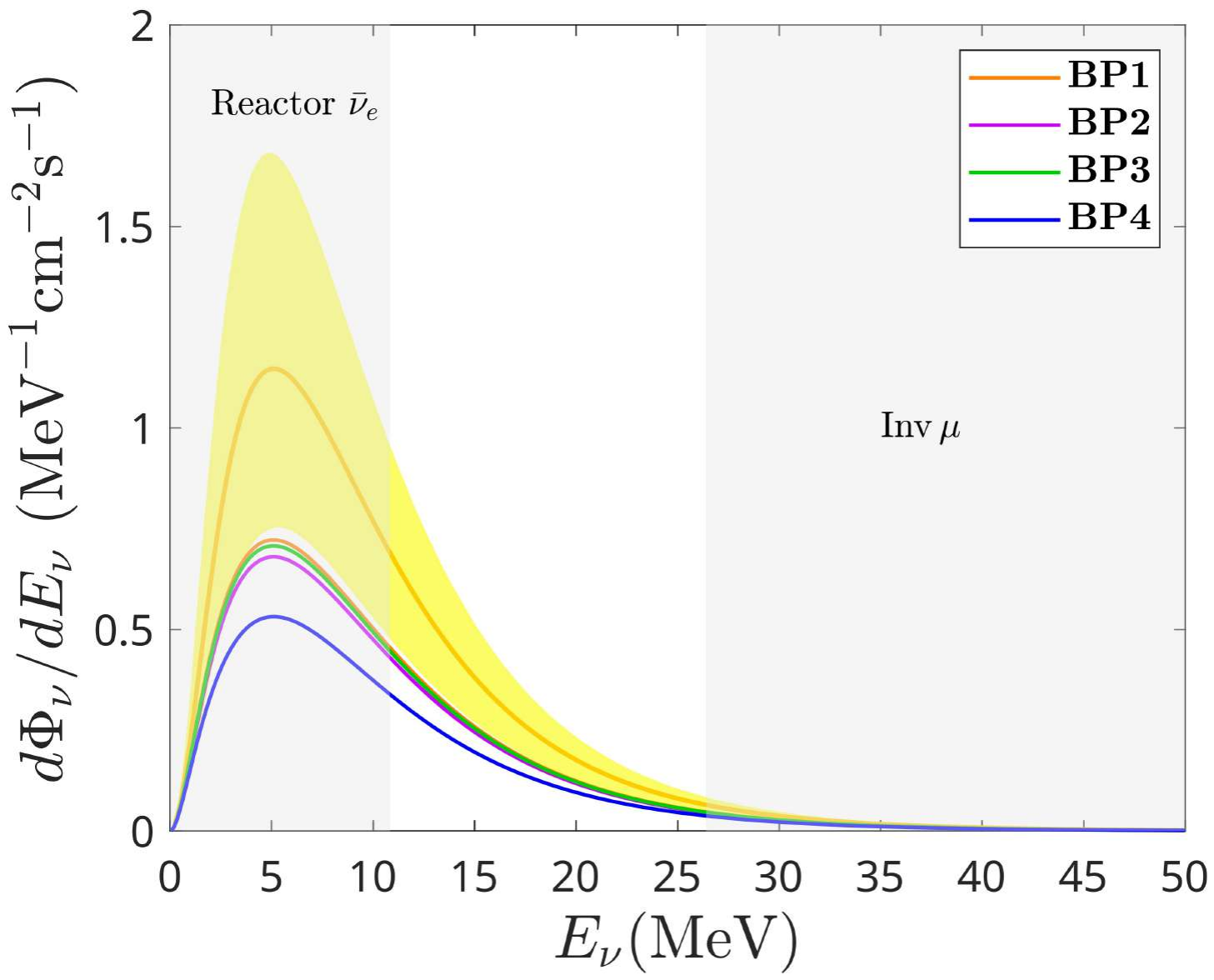}
    \caption{The DSNB flux for electron neutrino $\nu_e$ with temperature $T_\nu=6.6\,{\rm MeV}$ (dark yellow) and the fluxes of {\bf BP}s in Table \ref{table2} including DM attenuation. The yellow shaded region indicates the uncertainties arising from SFR.
	The gray shaded regions are the backgrounds of Reactor $\bar{\nu}_e$ and Inv$\,\mu$.}
    \label{bpdsnbflux}
\end{figure}

\noindent where $B=2^{1-\alpha/\beta}$, $C=2^{(\beta-\alpha)/\gamma}\cdot 5^{1-\beta/\gamma}$, $\dot{\rho}_0=0.0178^{+0.0035}_{-0.0036}\,{\rm M}_\odot\,{\rm yr}^{-1}\,{\rm Mpc}^{-3}$, and $\alpha=3.4\pm 0.2$, $\beta=-0.3\pm 0.2$, $\gamma=-3.5\pm 1$.
The initial mass function $\psi(M)$ is proportional to $M^{-2.35}$ \cite{Salpeter:1955it}. 
We demonstrate the flux of $\nu_e$ with $T_\nu=6.6\,{\rm MeV}$ 
in Fig.  \ref{bpdsnbflux} with dark yellow curve, and the yellow shaded region represents the uncertainties from SFR. 

To include the scattering of neutrinos and dark scalar $\phi$, we add the transmittance $T$ into Eq.~(\ref{DSNBflux}):
	\begin{equation}\label{dsnbflux}
	\frac{d\Phi_\nu}{dE_\nu}=\int_{0}^{z_{\rm max}}{\frac{R_{\rm CCSN}(z)F_\nu(E_\nu,z)T(E_\nu,z)}{H(z)}dz}.
	\end{equation}
We fixed $z_{\rm max}=5$ for which there is a reasonable amount of star formation. 
Numerically, we calculate the average DM column density at different redshift $z$, then run the $\nu$FATE over a energy range and get the flux attenuation at each $E_\nu$ and redshift $z$. Finally, we replace $T(E_\nu,z)$ in the above integrand by the flux attenuation from $\nu$FATE.

\begin{figure}
	\centering
    \includegraphics[width=0.49\linewidth]{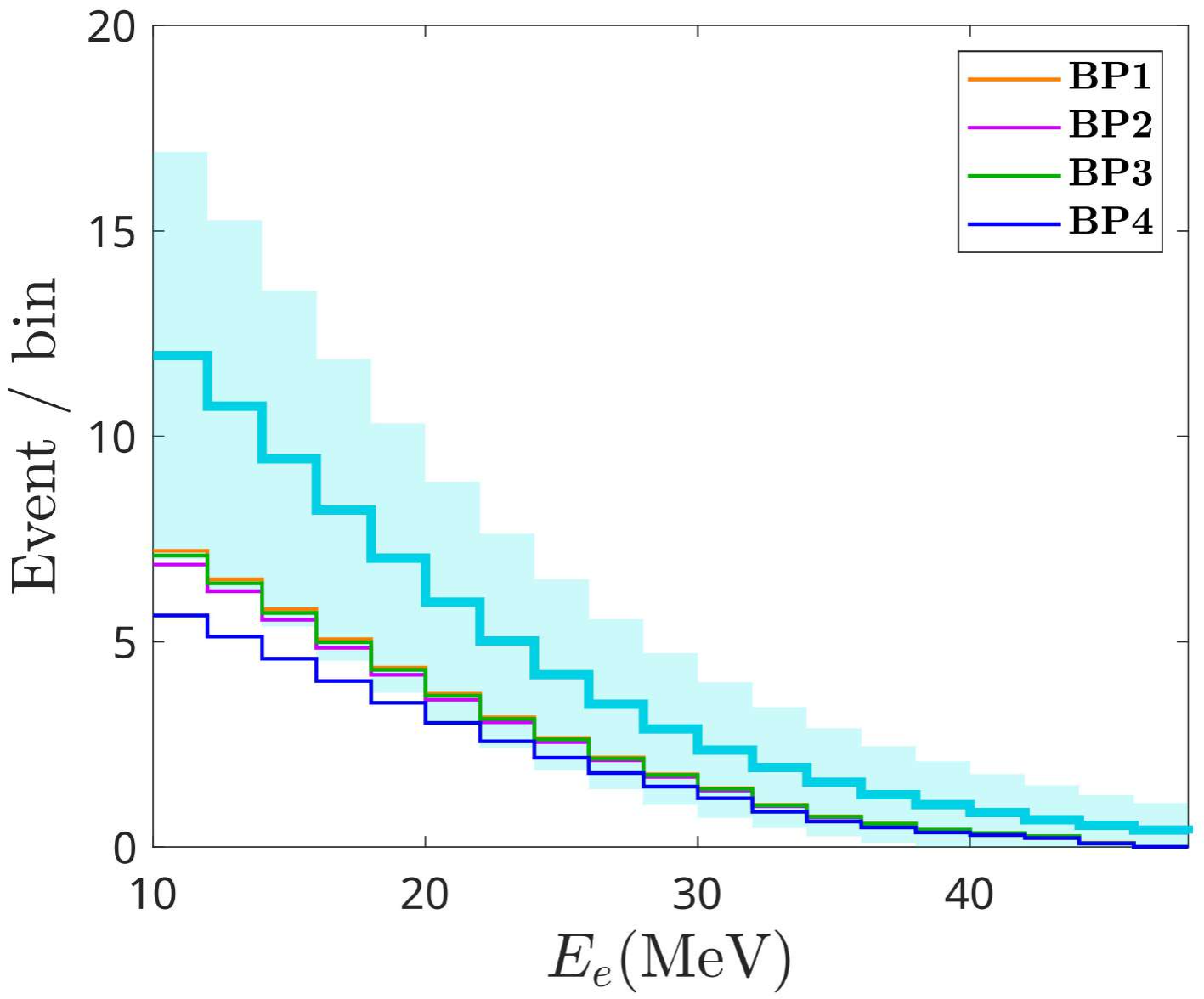}
    \includegraphics[width=0.49\linewidth]{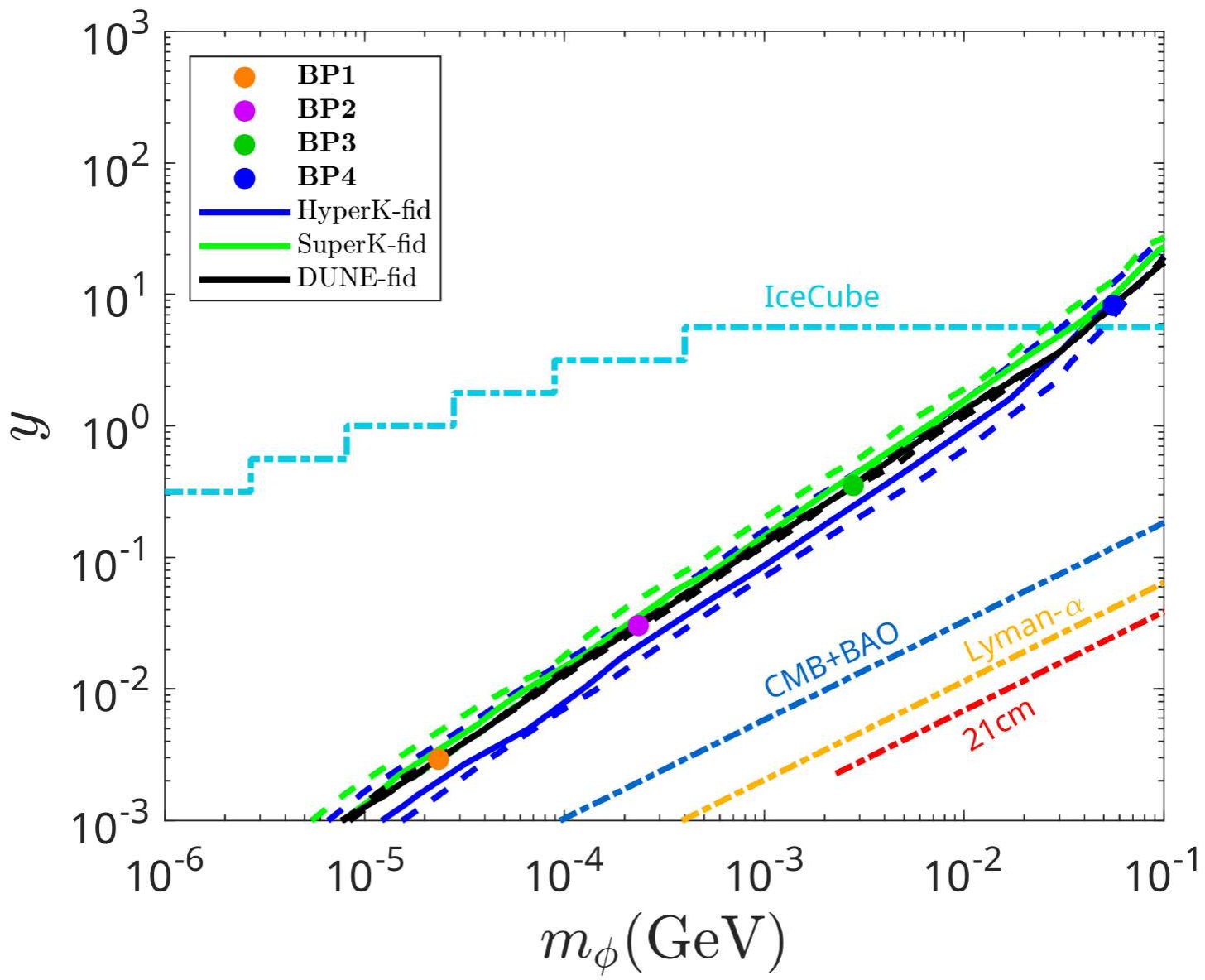}
	\caption{	
	(left) The event number for the DSNB flux with $T_\nu=6.6\,{\rm MeV}$ as a function of electron energy and the corresponding event numbers of benchmark points. 
    (right) Projecting sensitivities on $m_\phi$ and $y$, solid curves, obtained from the $\chi^2$ analysis, in which the corresponding dashed curves indicate the uncertainties from SFR parameters. The cyan stairs is the constraint from the IceCube observations \cite{Arguelles:2017atb}. The dash-dotted lines are the 90\% C.L. bounds of energy-independent DM-neutrino interactions~\cite{Zapata:2025huq}: CMB-BAO~\cite{Mosbech:2020ahp}, Lyman-$\alpha$~\cite{Hooper:2021rjc}, 21 cm~\cite{Dey:2022ini}. The corresponding values of {\bf BP}s are listed in Table~\ref{table2}.
	}
	\label{fig:ymphi2}
	\end{figure}

Detecting DSNB on Earth, 
the SK-IV data reveals mild excess above the background expectation~\cite{Super-Kamiokande:2021jaq}, which relies on the inverse beta decay, $\bar{\nu}_e + p \to e^+ + n$.
In the near future, we consider the HK~\cite{Hyper-Kamiokande:2022smq} and DUNE, where the later bases on the charged current interaction of liquid argon and low energy electron neutrino:
	\begin{equation}\label{argon}
	\nu_e+\prescript{40}{}{{\rm Ar}}\rightarrow e^-+\prescript{40}{}{{\rm K}}^\ast.
	\end{equation}
The neutrino incident energy can be written as
	\begin{table}
	\footnotesize
	\centering
	\begin{tabular}{c| c c c c}
		\hline
        &{\bf BP1}&{\bf BP2}&{\bf BP3}&{\bf BP4}\\
        \hline
        $m_\phi/{\rm GeV}$ &$2.34\times10^{-5}$&$2.34\times10^{-4}$&$2.79\times10^{-3}$&$5.57\times10^{-2}$\\        $y$&$2.92\times10^{-3}$&$3.04\times10^{-2}$&$3.53\times10^{-1}$&$8.26$\\
        \hline
	\end{tabular}
	\caption{Parameter values of {\bf BP}s in Fig.~\ref{bpdsnbflux} and Fig.~\ref{fig:ymphi2}.}
	\label{table2}
\end{table}
    \begin{equation}
	E_\nu=E_e+[(m^{\rm g}_{\rm K}+E_x)-m^{\rm g}_{\rm Ar}]+T_{\rm K},
	\end{equation}
where $E_e$ is the energy of outgoing electron, $m^{\rm g}_{\rm K}\,(m^{\rm g}_{\rm Ar})$ is the ground-state mass of potassium (argon), $E_x$ is the excitation energy, and $T_{\rm K}$ is the recoil kinetic energy of ${\rm K}$. For low energy neutrinos, one may neglect $T_{\rm K}$. The mass difference $m^{\rm g}_{\rm K}-m^{\rm g}_{\rm Ar}$ is around $1.505\,{\rm MeV}$. 
The total $\nu$-Ar cross section $\sigma_{\nu {\rm Ar}}$ of this interaction in the CM frame is given in Appendix~\ref{appendix:B}, where we use the data set of the nuclear matrix elements from Ref.~\cite{Gardiner:2018zfg}. 
By assuming 400 kton-years (3.8 Mton-years) of exposure of a DUNE (HK) detector, we may calculate the numbers of event of the scattering Eq.~(\ref{argon}). The event number is given by
	\begin{equation}\label{eventnumber}
	N^{\rm DSNB}(E_e)=\epsilon N_{\rm Ar}\int{dE_\nu \frac{d\Phi_\nu}{dE_\nu}\sigma_{\nu {\rm Ar}}},~{\rm or}~~
    \epsilon N_{\rm p}\int{dE_\nu \frac{d\Phi_\nu}{dE_\nu}\sigma_{\nu {\rm p}}}
	\end{equation}
For DUNE, the $\epsilon$ is the detector efficiency and is assumed to be {\color{blue}$86\%$}, $N_{\rm Ar}$ is the number of target argon~\cite{Moller:2018kpn}. Meanwhile, for SK/HK, we use the energy-dependent efficiency according to Ref.~\cite{Super-Kamiokande:2021jaq}.
Taking DUNE for instance, the binned event number of the DSNB flux is shown in the left panel of Fig.~\ref{fig:ymphi2}. 
The uncertainty including the systematic and statistic errors (blue shaded region) of each bin $i$ is determined by 
	\begin{equation}
	\sigma^{\rm tot}_i=\sqrt{(\sigma^{\rm sys}_i)^2+N_i},~~~~\text{where}~~\sigma^{\rm sys}_i=\frac{N^{+}_i-N^{-}_i}{2}.
	\end{equation}
$N_i$, $N^{+}_i$ and $N^{-}_i$ are the event numbers for the DSNB fluxes with fiducial, upper, and lower SFR parameters. 

From Eq.~(\ref{dsnbflux}) we can calculate the event number of the scattered DSNB flux and recast to constraint on $m_\phi$ and $y$ by using the Poisson likelihood 
	\begin{equation}\label{chisquare}
	\chi^2= 2\sum_i\left( N_i^{\rm exp}-N_i^{\rm obs}+N_i^{\rm obs} \ln \frac{N_i^{\rm obs}}{N_i^{\rm exp}} \right),
	\end{equation} 
where $N_i^{\rm obs}$ is the observed number of events of the $i$-th bin and $N_i^{\rm exp}=N_i^{\rm bkgd}+N_i^{\rm DSNB}$ including the background and
the DSNB event numbers computed from Eq.~(\ref{eventnumber}) and Eq.~(\ref{dsnbflux}). Notice that the background event originating from SK analysis~\cite{Super-Kamiokande:2021jaq} is rescaled to the HK exposure.
For DUNE, we refer to background event from the atmospheric charge current estimated in Ref.~\cite{Moller:2018kpn}, and assume that the hypothetical observed number of event is consistent with zero $\nu\phi$ interaction.
To obtain the conservative bounds, we adopted the 
fiducial, upper, and lower SFR parameters values to demonstrate the uncertainty from CCSN calculation, and scanned over the parameter space
    \begin{equation}
    10^{-6}\leq m_\phi/{\rm GeV}\leq0.1,\quad 10^{-3}\leq y\leq 10^3
    \end{equation}
to calculate transmittance $T$.
Due to the event excess from SK, the DSNB flux without $\nu\phi$ scattering is favored by the data, thus leads to lower value of $\chi^2_{\rm no-\nu\phi}$. Subsequently, including the DM attenuation enlarges $\chi^2_{\nu\phi}$.
We fixed $m_F/m_\phi=1.1$ in our analysis, and
the sensitivity is shown in the right panel of Fig.~\ref{fig:ymphi2} by the solid curves, where the above regions are disfavored by $2\sigma$ (i.e. $\chi^2_{\nu\phi}-\chi^2_{\rm no-\nu\phi}\geq 4$), and the uncertainties of SFR parameter are depicted by the corresponding dashed curves.
In Fig.~\ref{fig:ymphi2}, the $y$ value increases as $m_\phi$ increases, this is because the cross section is inversely proportional to the DM mass and mediator mass according to Eq.~(\ref{limit1}) and (\ref{limit3}). Since $\sigma_{\nu\phi}\propto m^{-1}_\phi$ in $m_\phi/{\rm GeV}\lesssim 10^{-3}$ and $\sigma_{\nu\phi}\propto m^{-4}_\phi$ in $m_\phi/{\rm GeV}\gtrsim 10^{-1}$, the slope of the margin is slightly increased.
There are four selected benchmark points on the margin of DUNE fiducial curve. The event numbers of each {\bf BP} were significantly attenuated at low electron recoil energy. 
The corresponding DSNB fluxes of {\bf BP}s are also shown in Fig. \ref{bpdsnbflux}. There is an open energy window from 10.8 MeV to 26.4 MeV which is sandwiched by the overwhelming backgrounds of reactor $\bar{\nu}_e$ and inverse $\mu$ decay. We can see that all of the {\bf BP}s can be distinguished from the unattenuated DSNB flux and produce the detectable suppression signal.


\bigskip

\section{Active Galactic Nuclei NGC 1068 and TXS 0506+056}\label{agn}

AGNs of NGC 1068 and TXS 0506+056 produce ultra high-energy neutrinos have been detected at IceCube~\cite{IceCube:2022der,IceCube:2018cha}. 
It was suggested that the DM density around the SMBH at the center of an AGN, for $r \leq R_{\rm sp}$, 
may form a spike, 
which is given by~\cite{Herrera:2023nww}
	\begin{equation}\label{rhosp}
	\rho_{\rm sp}=\rho_R g_\gamma(r) \left(\frac{R_{\rm sp}}{r}\right)^{\gamma_{\rm sp}},
	\end{equation}
where \cite{Cline:2023tkp}
	\begin{equation}
	R_{\rm sp}=\left(\frac{M_{\rm BH}}{4\pi \rho_0 r_0[f(r_h)-f(r_i)]}\right)^{3/4}
	\end{equation}
represents the size of the spike, $r_h$ is the influence radius of supermassive black hole~\cite{Gnedin:2003rj}, and $r_i$ is the inner radius of the spike here we take it to be four times the Schwarzschild radius, i.e. $r_i=4R_{\rm s}$~\cite{Herrera:2023nww}. 
\begin{table}
	\footnotesize
	\centering
	\begin{tabular}{c| c c c c}
		\hline
        &{\bf NGC BP1}&{\bf NGC BP2}&{\bf NGC BP3}&{\bf NGC BP4}\\
        \hline
        $m_\phi/{\rm GeV}$ &$2.63\times10^{-5}$&$8.94\times10^{-4}$&$1.81\times10^{-2}$&$5.34\times10^{-1}$\\  
        $\sigma_0/{\rm cm}^2$&$7.15\times10^{-37}$&$2.90\times10^{-34}$&$2.84\times10^{-31}$&$7.73\times10^{-28}$\\
        $\langle\sigma_{\phi\phi} v\rangle/{\rm cm}^2$ &$2.50\times10^{-51}$&$3.45\times10^{-47}$&$6.85\times10^{-43}$&$5.50\times10^{-38}$\\
        \hline
	\end{tabular}
	\caption{Parameter values of {\bf NGC BP}s in Fig.~\ref{fig:agnconstraint}.}
	\label{table_AGN}
\end{table}
In addition, we take the DM self annihilation into account, the density Eq.~(\ref{rhosp}) should be modified by
	\begin{equation}
	\frac{\rho_{\rm sp}(r)\rho_{\rm sat}}{\rho_{\rm sp}(r)+\rho_{\rm sat}}
	\end{equation}
which reaches the saturation density (refer Appendix~\ref{appendix:C} for more details)
	\begin{equation}
	\rho_{\rm sat}=\frac{m_\phi}{\langle\sigma_{\phi\phi} v\rangle t_{\rm BH}}\,,
	\end{equation}
where $v \sim 10^{-3}\,c$ is the thermal average velocity of the DM, 
$t_{\rm BH}$ is the age of the central black hole.
We adopt nonself conjugate $\phi\phi^*$ annihilation cross section $\sigma_{\phi\phi}$, 
and their expression is included in Appendix~\ref{app:annihilation}.
For $r>R_{\rm sp}$, it recovers the Navarro-Frenk-White (NFW) profile
	\begin{equation}
	\rho_{\rm NFW}=\rho_0\left(\dfrac{r}{r_0}\right)^{-\gamma}\left(1+\dfrac{r}{r_0}\right)^{-(3-\gamma)}.
	\end{equation}
Taking $\gamma=1$, we have $\gamma_{\rm sp}=\frac{9-2\gamma}{4-\gamma}=\frac{7}{3}$, $g_\gamma(r)\simeq (1-\frac{4R_s}{r})^3$, and
	\begin{equation}
	f(r)=r^{-\gamma_{\rm sp}}\left(\frac{r^3}{3-\gamma_{\rm sp}}+\frac{12R_s r^2}{\gamma_{\rm sp}-2}-\frac{48 R^2_s r}{\gamma_{\rm sp}-1}+\frac{64 R^3_s}{\gamma_{\rm sp}}\right)\,.
	\end{equation}
The DM density profile then follows \cite{Cline:2023tkp}
	\begin{equation}\label{spike}
	\rho_{\rm DM}(r)=
	\begin{cases}
	0&, r\leq 4 R_s,\\
	\dfrac{\rho_{\rm sp}(r)\rho_{\rm sat}}{\rho_{\rm sp}(r)+\rho_{\rm sat}}&,4R_s\leq r\leq R_{\rm sp},\\[2ex]
	\dfrac{\rho_{\rm NFW}(r)\rho_{\rm sat}}{\rho_{\rm NFW}(r)+\rho_{\rm sat}}&, r\geq R_{\rm sp}.
	\end{cases}
	\end{equation}

Since the distance between NGC 1068 and the Earth is around $14.4\,{\rm Mpc}$, the effect of redshift is negligible.
The estimates of NGC 1068 supermassive black hole vary, and we adopt mass $M_{\rm BH}\simeq 10^7\, M_\odot$, age $t_{\rm BH}\simeq 10^9$ yrs, and influence radius $r_h=6.5\times 10^5 R_{\rm s}$~\cite{Cline:2023tkp}.
The optical depth of the emitted high-energy neutrinos is obtained via integrating the DM number density along the line of sight from $r=4 R_{\rm s}$
	\begin{equation}\label{tau}
	\tau=\sigma_{\nu\phi}\int^{14.4\,{\rm Mpc}}_{4R_s}{\frac{\rho_{\rm DM}(r)}{m_\phi}dr}.
	\end{equation} 
The NGC 1068 produces neutrinos with an energy interval $[E_{\rm min},E_{\rm max}]=[1.5\,{\rm TeV},15\,{\rm TeV}]$, which is much larger than the mass range of $\phi$ that we considered for DSNB \sout{and DUNE}.
For AGNs emitting high-energy neutrinos, we adopt the kinematic region and parametrize cross section from Eq.~(\ref{parametrized}), which is linear in $E_\nu$, to compute the optical depth.
Then Eq.(\ref{tau}) becomes
	\begin{equation}\label{eq:tau}
	\tau=\sigma_{\nu\phi}\int^{14.4\,{\rm Mpc}}_{4R_s}{\frac{\rho_{\rm DM}(r)}{m_\phi}dr}=\frac{y^4E_\nu}{32\pi m^4_F}\int^{14.4\,{\rm Mpc}}_{4R_s}{\rho_{\rm DM}(r)dr}.
	\end{equation}

\begin{figure}
	\centering
	\includegraphics[width=0.495\linewidth]{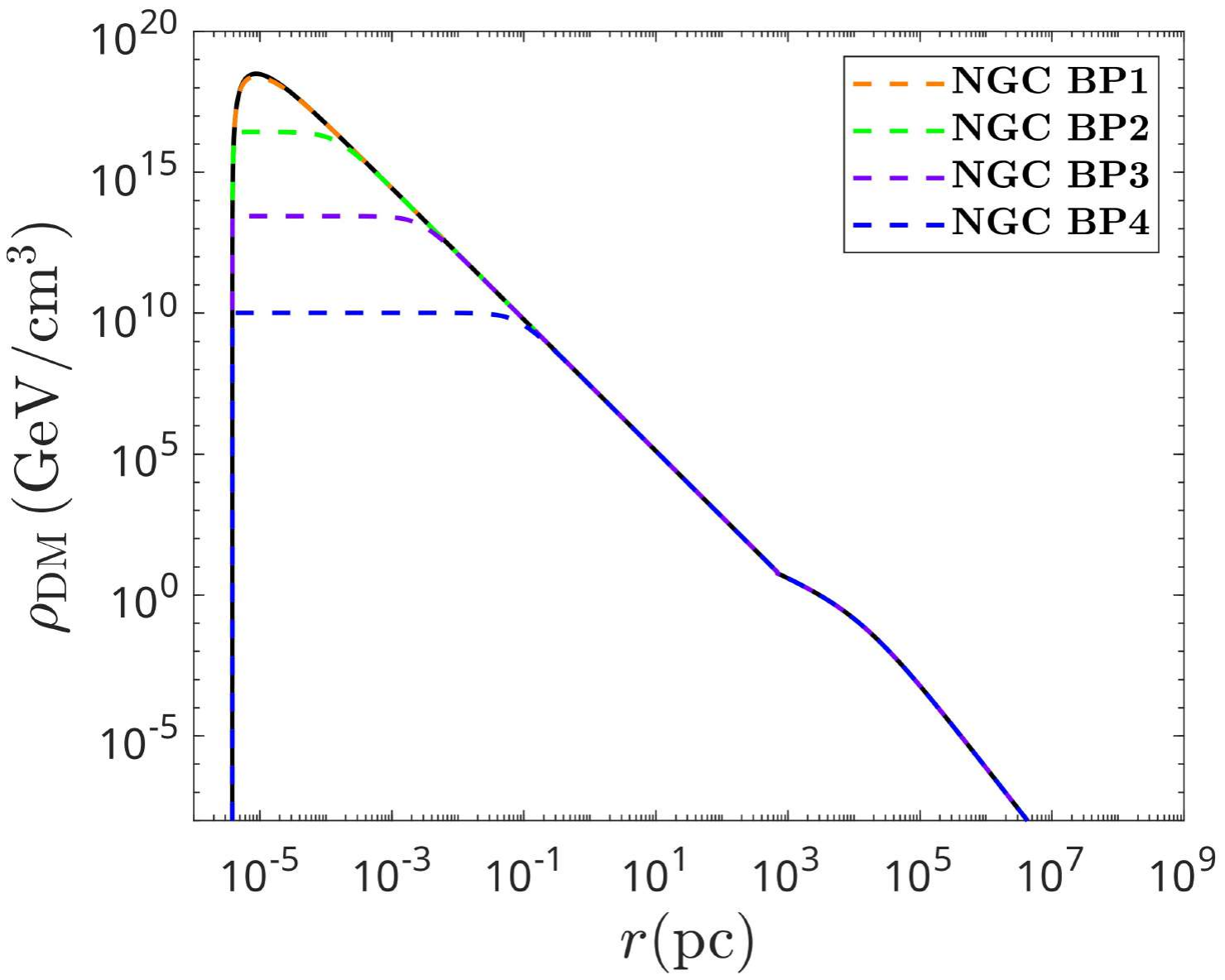}
	\includegraphics[width=0.495\linewidth]{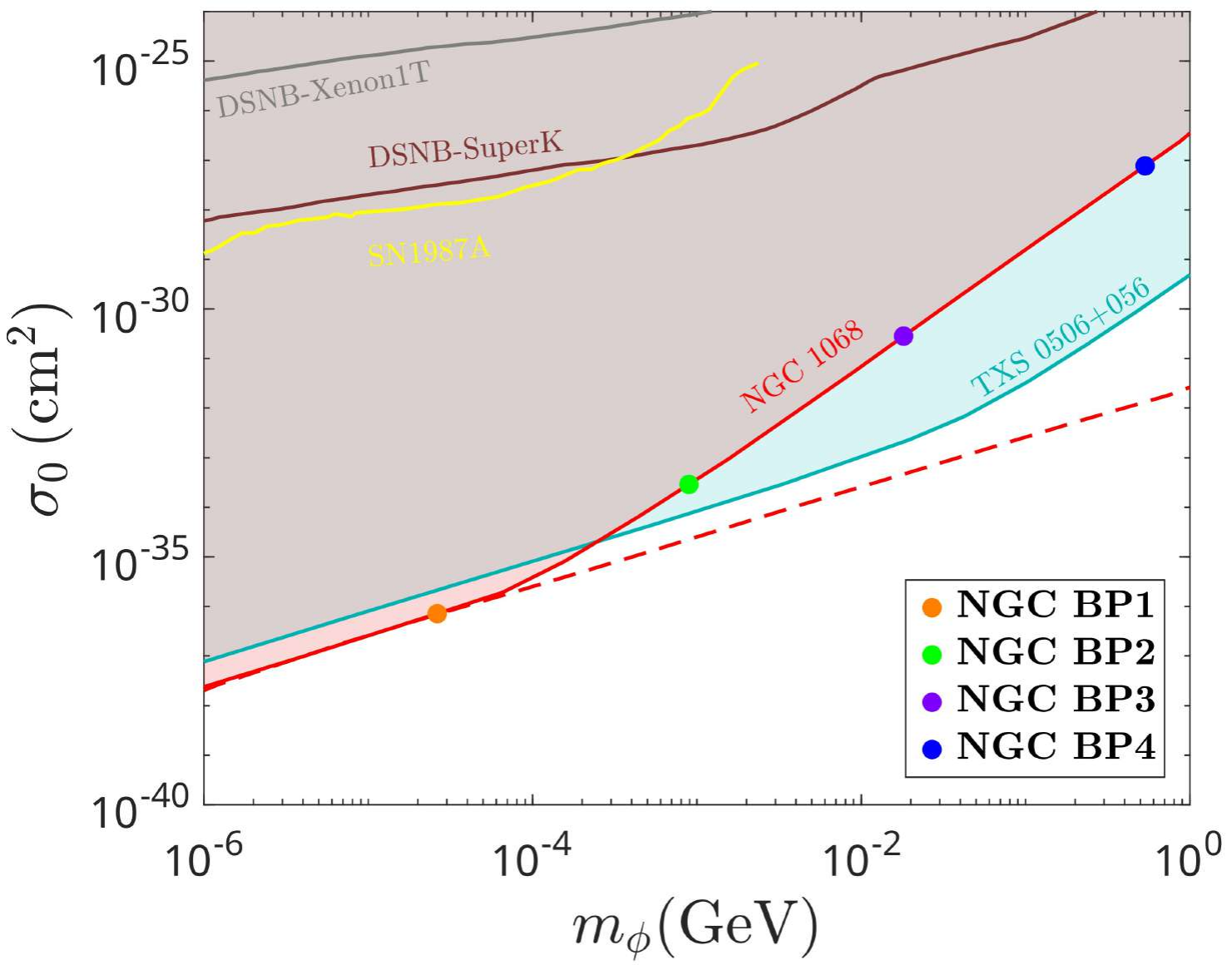}
	\caption{(left) The DM density profile of NGC 1068 with $\rho_0=0.35\,{\rm GeV}/{\rm cm}^3$, $M_{\rm BH}=10^7 M_{\odot}$, $r_0=13\,{\rm kpc}$, $t_{\rm BH}=10^{9}\,{\rm years}$, $r_h=6.5\times10^5R_s$, $r_i=4R_s$. The black-solid curve depicts the density without $\phi\phi^*$ annihilation;
		the dashed curves are the modified densities of the {\bf NGC BP}s'. The corresponding values of {\bf NGC BP}s are listed in Table~\ref{table_AGN}. (right) The constraints on $\sigma_0$ defined in Eq.~(\ref{parametrized}) by requiring $Q=0.5$  ($Q=0.05$) for NGC 1068 (TXS 0506+056), $m_F=10\,{\rm TeV}$, and $E_0=10\,{\rm TeV}$.
		The solid-red curve is the constraint for NGC 1068, and the cyan for TXS 0506+056. 
        The dashed line represents the constraint from NGC 1068 without $\phi\phi^*$ annihilation. 
        For comparison, we also include the limits from DSNB-Xenon1T/SN1987A~\cite{Cline:2023tkp,Lin:2022dbl} and DSNB-SuperK~\cite{Ghosh:2021vkt}, however they imposed more stringent assumptions on DM-neutrino and DM-electron interactions.
	}
	\label{fig:agnconstraint}
\end{figure}

The NGC 1068 neutrino flux is measured by the IceCube collaboration in terms of power law spectrum
$\Phi(E_\nu)\propto (E_\nu)^{-\hat{\gamma}}$ with the best fit value $\hat{\gamma}=3.2$~\cite{IceCube:2022der}.
Assuming no DM attenuation, we associate the $N$, measured events at IceCube, with NGC 1068 neutrino flux via~\cite{Doring:2023vmk}
	\begin{equation}\label{icecubenumber}
	N=t\int^{E_{\rm max}}_{E_{\rm min}}{A_{\rm eff}(E_\nu)\Phi(E_\nu)dE_\nu},
	\end{equation}
where $t$ is the exposure time, $A_{\rm eff}$ is the effective area of the IceCube detector \cite{Cline:2023tkp}.
To compare with the event numbers influenced by $\nu \phi$ scattering and to give the constraints on $m_\phi$ and $\sigma_0$, 
we require the following inequality:
	\begin{equation}\label{ratio}
	\frac{N_{\rm sct}}{N}=\frac{\int^{E_{\rm max}}_{E_{\rm min}}{A_{\rm eff}(E_\nu)\Phi(E_\nu)e^{-\tau(E_\nu)} dE_\nu}}{\int^{E_{\rm max}}_{E_{\rm min}}{A_{\rm eff}(E_\nu)\Phi(E_\nu)dE_\nu}}\geq Q.
	\end{equation}
$N_{\rm sct}$ is the event number with $\nu\phi$ scattering included. 
More specifically, we implement the DM column density from Eq.(\ref{tau}) into $\nu$FATE, then replace the $e^{-\tau}$ in the above expression by the attenuation factor calculated from $\nu$FATE.
Including the uncertainties from IceCube~\cite{IceCube:2022der,Cline:2022qld}, we set $Q=0.5$ ($Q=0.05$) for NGC 1068 (TXS 0506+056) and the corresponding constraint is shown in the right panel of Fig.~\ref{fig:agnconstraint}.\footnote{Caveat: Our constraints depend on the assumption of the AGN flux models. However, these models are posterior calculations that aim to explain the IceCube observations, which does not mean we fully understand the mechanisms of NGC 1068 nor TXS 0506+056.}
Four benchmark points are chosen on the NGC constraint, and the corresponding DM density profiles are shown in the left panel of Fig.~\ref{fig:agnconstraint}. In right panel of Fig.~\ref{fig:agnconstraint}, the solid curves for NGC 1068 and TXS 0506+056 represent the constraints which assume that the $\nu\phi$ scattering and $\phi\phi^*$ annihilation cross sections are correlated from the same interaction of Eq.~(\ref{newinteraction}), while the red-dashed curve ignores the $\phi\phi^*$ annihilation cross section thus there is no spike density suppression from saturation effect. It shows that when $m_\phi$ is light enough, i.e. {\bf NGC BP1}, the DM annihilation is negligible. Whereas, when $m_\phi$ is larger than $0.1\,{\rm MeV}$, i.e. {\bf NGC BP2} to {\bf NGC BP4}, the $\phi\phi^\ast$ annihilation cross section $\sigma_{\phi\phi}$ becomes significant to alleviate the saturation density, hence the $\sigma_0$ must increases to compensate the deficit in the DM density, which causes the raising of solid-red curve in right panel of Fig.~\ref{fig:agnconstraint}.

Another interesting AGN is the TXS 0506+056. It is a much more distant neutrino source than NGC 1068. Its redshift is measured as $z=0.336$~\cite{Paiano:2018qeq}  
and corresponding distance is around $1.37\,{\rm Gpc}$ away from Earth. The relevant parameters for TXS 0506+056 are given by 
$r_i=4R_s$, $r_h\simeq 10^5R_s$, $M_{\rm BH}\simeq 3.09\times10^8 M_\odot$, and $t_{\rm BH}=10^9\,{\rm years}$~\cite{Cline:2022qld}. 
For $\rho_0$ and $r_0$, they can be computed by $\rho_0=0.154\,{\rm GeV}/{\rm cm}^3$, $r_0=42.36\,{\rm kpc}$  (see Appendix~\ref{appendix:C} for more details).
Because of the cosmological distance of TXS 0506+056,
we need to include the effect of cosmological expansion. 
From Eq.~(\ref{taured}) and (\ref{eq:tau}) we have,
	\begin{align}\label{tauTXS}
	\tau(E_\nu,0.336)
	=\frac{y^4E_\nu}{32\pi m^4_F}\int^{0.336}_0{\frac{\rho_{\rm DM}(d_0-d(z))(1+z)^3}{H(z)}dz},
	\end{align}
where $d(z)$ is the comoving distance
	\begin{equation}
	d(z)=\int^z_0{\frac{dz^\prime}{H(z^\prime)}}
	\end{equation}
and $d_0=1.37\,{\rm Gpc}$.
The main contribution of DM density is within the $R_{\rm sp}\ (\sim 3.1\,{\rm kpc})$, which is much smaller than $d_0$. The redshifts for the density with $r\lsim 1\,{\rm Mpc}$ are nearly a constant, hence we approximate Eq.~(\ref{tauTXS}) by 
	\begin{equation}
	\tau(E_\nu,0.336)\simeq \frac{y^4 E_\nu}{32\pi m^4_F}(1.336)^3\int_{4R_s}^{R}\rho_{\rm DM}dr,
	\end{equation}
where the upper limit $R$ is taken to be $R=2\,{\rm Mpc}$, and $\rho_{\rm DM}(R)\simeq 1\times 10^{-6}\,{\rm GeV}/{\rm cm}^3$ is consistent with the average DM density in the extragalactic medium.
The contribution for $r > R$ is negligible.

The effective area for IceCube can be parameterized as in Ref.~\cite{Cline:2022qld},  
and the best fit power of neutrino flux for TXS 0506+056 is $\hat{\gamma}=2$~\cite{Doring:2023vmk}. 
The energy range of the neutrinos produced from TXS 0506+056 is $[E_{\rm min},E_{\rm max}]=[40\,{\rm TeV},4000\,{\rm TeV}]$. 
We calculate the constraint on $\sigma_0$ for TXS 0506+056 according to Eq.(\ref{icecubenumber}) and Eq.(\ref{ratio}).
Fig. \ref{fig:agnconstraint} shows that the constraint for TXS 0506+056 is slightly stronger than that of NGC 1068  even though the $Q_{\rm TXS}$ ($=0.05$) is much lower than $Q_{\rm NGC}$ ($=0.5$), this is because cross section $\sigma_{\nu\phi}$ is proportional to $E_\nu$, 
and the neutrino energy is much higher of TXS 0506+056. 
Therefore, the coupling constant $y$ must be smaller to prevent the intense scattering. 
The $\phi\phi^*$ annihilation has mild contribution in this case when $m_\phi$ is larger than 50 MeV.

\begin{figure}
	\centering
	\includegraphics[width=0.496\linewidth]{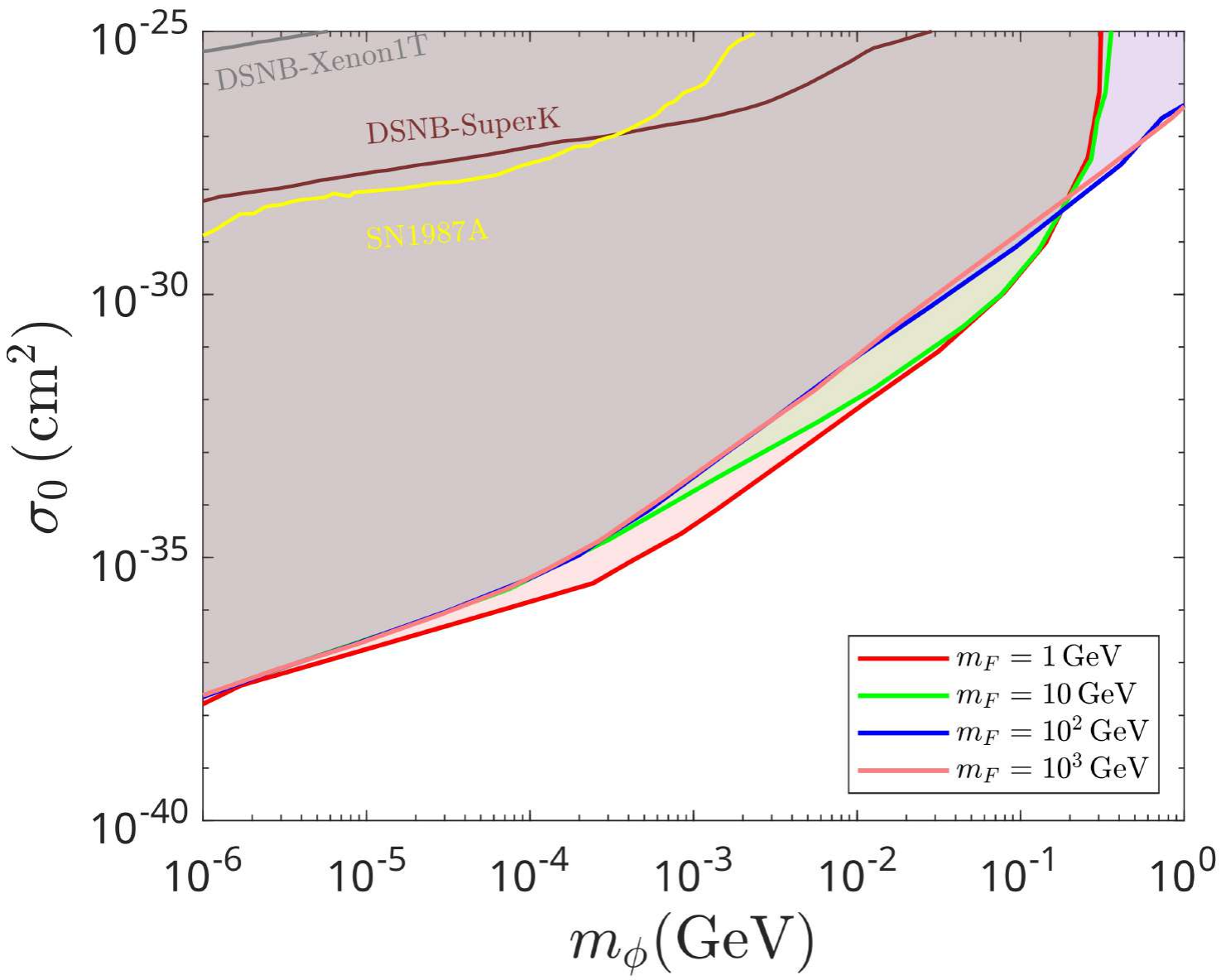}
	\includegraphics[width=0.496\linewidth]{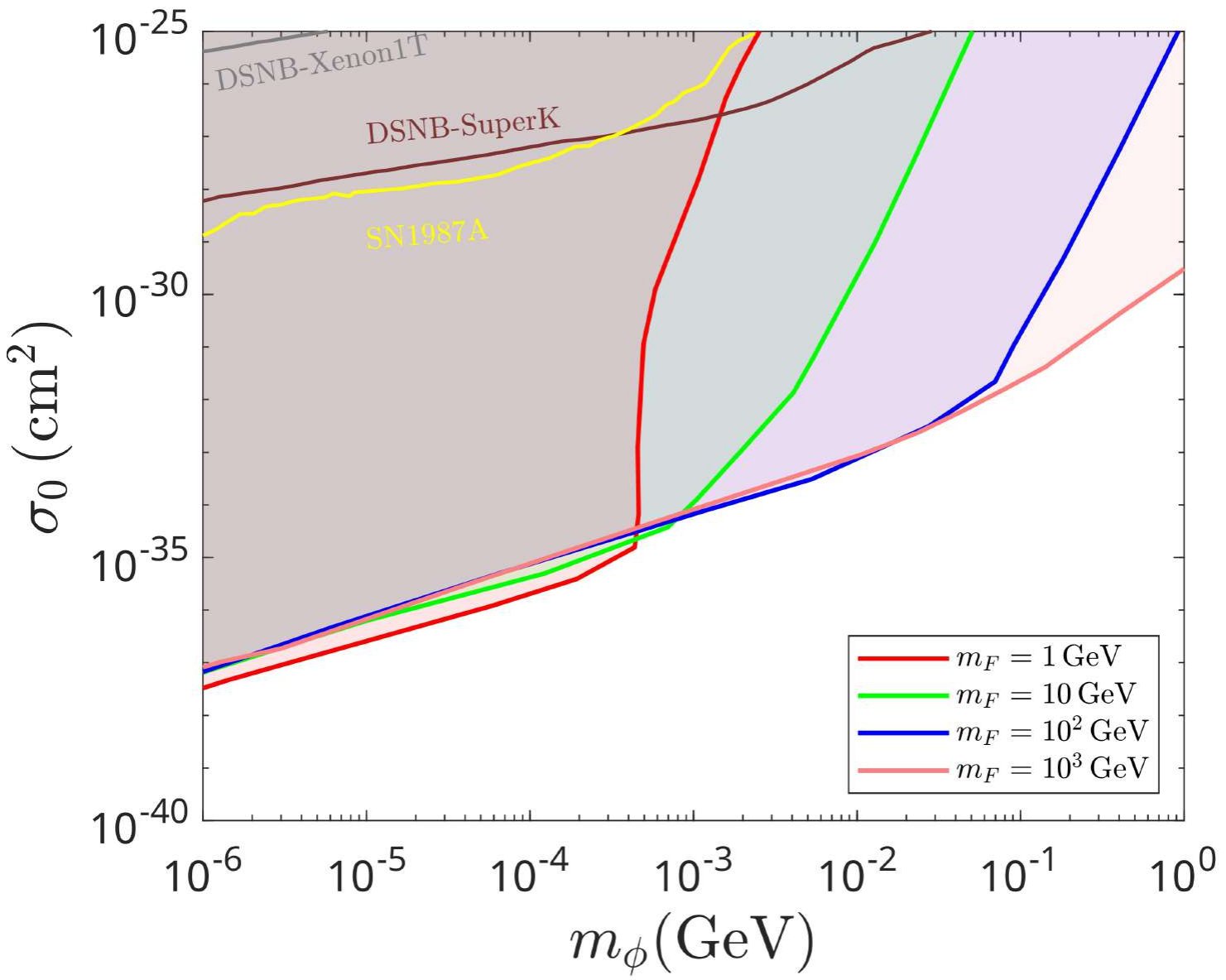}
	\caption{The constraints on $\sigma_0$ defined in Eq.(\ref{parametrized}) with $Q=0.5$ ($Q=0.05$) for NGC 1068 (TXS 0506+056) and $E_0=10\,{\rm TeV}$, varying $m_F=1\,{\rm GeV},\, 10\,{\rm GeV},\,100\,{\rm GeV},\, 1000\,{\rm GeV}$. Left (Right) panel is for NGC 1068 (TXS 0506+056).}
	\label{fig:sigma0mphi}
\end{figure}

We also calculated the constraints of $\sigma_0$ without assuming $m_F\gg m_\phi$ for comparison. In this case, we use the exact scattering cross section Eq.~(\ref{nuphicrosssection2}) and consider $m_F$ from $1\,{\rm GeV}$ to $1000\,{\rm GeV}$.
In Fig.~\ref{fig:sigma0mphi}, we plot $\sigma_0$ against $m_\phi$ with various $m_{F}$ and compare the results in Fig.~\ref{fig:agnconstraint} which fixes $m_{F}=10\,{\rm TeV}$.
For NGC 1068, the behaviors of these curves at low $m_\phi$ ($m_F\gg m_\phi$ still holds) are the same; the DM annihilation is irrelevant and the parameterized cross section Eq.~(\ref{parametrized}) is still valid.
As $m_\phi$ gets larger, the annihilation starts to contribute. When $m_\phi$ is large enough, such as $m_\phi E_\nu>m^2_F$ in the denominator of Eq.~(\ref{nuphicrosssection3}), the scattering cross section is proportional to $E^{-1}_\nu$, 
and thus $y$ must be drastically increased to satisfy $N_{\rm sct}/N\leq 0.5$, which is shown in Fig.~\ref{fig:sigma0mphi}. 
For TXS 0506+056, 
%
%
, since the energy range of TXS-emitted neutrinos is much larger than that of NGC, the condition $m_\phi E_\nu > m^2_F$ and the $\sigma_{\nu\phi}$ converting into $E_\nu^{-1}$ dependent occur at lighter $m_\phi$ comparing to the NGC one in the left panel.
Due to the same reason, when $m_F$ is getting lighter,
the upper bounds of $\sigma_0$ of TXS deviate faster from the one with $m_F=10{\rm TeV}$ than those of NGC do.

\bigskip

\section{Conclusion}\label{conclusion}
We utilize two energy-distinctive astrophysical neutrino sources, DSNB and AGN, to constrain on the neutrino-DM interaction which is induced via exchanging a fermionic mediator. Given this, the neutrino-DM cross section exhibits different energy dependence in various kinematic parameter regions.
Considering the attenuation on the neutrino flux during propagation, we derive the upper bounds of the coupling constant $y$ and scattering cross section $\sigma_{\nu \phi}$.
%
For $\mathcal{O}(10\,{\rm MeV})$ neutrino from DSNB,
we calculate the optical depth and estimate the event numbers at DUNE detector through $\nu{\rm Ar}$ scattering, then perform the chi-square test to find the upper bound of $y$ as function of $m_\phi$.
Fig.~\ref{fig:ymphi2} shows that the upper bound of $y$ 
is small enough for the parameter region $E_\nu \gg m_\phi \simeq m_F$ and $\sigma_{\nu \phi} \propto E^{-1}_\nu$, so that the amplitude satisfies perturbativity condition.
Meanwhile, within the energy range $10.8\leq E_\nu/{\rm MeV}\leq 26.4$, the benchmark points in Table \ref{table2} predict the detectable attenuation of DSNB fluxes.

AGNs, NGC 1068 and TXS 0506+056, are the second sources we considered.
The energy of neutrinos emitted from these two AGNs are much higher than those of DSNB, 
we thus focus on the kinematic region, $m^2_F \gg E_\nu m_\phi \gg m^2_\phi$, such that $\sigma_{\nu\phi} \propto E_\nu$.
Including the spike DM density profile around AGN supermassive black hole, 
the $\nu\phi$ scattering and the $\phi\phi^\ast$ annihilation cross sections both modify the neutrino flux at IceCube detector.
Fig.~\ref{fig:agnconstraint} shows that both NGC 1068 and TXS 0506+056 can provide more stringent constraints than the DSNB-Xenon1T and SuperK.
For NGC 1068, due to the fact that the DM annihilation becomes significant enough to suppress the saturation density when $m_\phi/{\rm GeV}\gsim 10^{-4}$ the slop of $\sigma_0$ upper bound grows steeper.
In particular, $\sigma_0$ takes the value from $2.4\times10^{-38}\,{\rm cm}^2$ to $3.3\times10^{-36}\,{\rm cm}^2$ for $10^{-6}\leq m_\phi/{\rm GeV}\leq 10^{-4}$ and grows from $3.3\times10^{-36}\,{\rm cm}^2$ to $4.2\times10^{-27}\,{\rm cm}^2$ for $10^{-4}\leq m_\phi/{\rm GeV}\leq 1$ in which the annihilation starts influence.
Conversely, the annihilation is negligible for TXS 0506+056 when $10^{-6}\leq m_\phi/{\rm GeV}\leq 5\times10^{-2}$ due to its extremely high-energy neutrinos, 
since the coupling constant $y$ must be reduced to compensate the intense $\nu\phi$ scattering.
As a result, $\sigma_0$ maintains the linearity from $7.2\times10^{-38}\,{\rm cm}^2$ to $6.8\times10^{-33}\,{\rm cm}^2$ for $10^{-6}\leq m_\phi/{\rm GeV}\leq 5\times10^{-2}$ but grows from $6.8\times10^{-33}\,{\rm cm}^2$ to $5.7\times10^{-30}\,{\rm cm}^2$ for $5\times 10^{-2}\leq m_\phi/{\rm GeV}\leq 1$.
If we decrease the value of $m_F$, as shown in Fig.~\ref{fig:sigma0mphi}, $\sigma_0$ increases when the condition $m_F\gg m_\phi$ breaks down. 
This is the direct consequence of the inverse proportionality between the exact cross section Eq.(\ref{nuphicrosssection3}) and the neutrino energy, i.e. $\sigma_{\nu\phi} \propto 1/E_\nu$.
For TXS 0506+056,
with $m_{F}=1\,{\rm GeV}$ and $m_\phi\gtrsim 1.5\,{\rm MeV}$, the upper bound is getting even weaker than DSNB-SuperK. Final remark is that our results, corresponding to AGNs, heavily rely on the existence of DM spike at the center of AGN. The formation of DM spike near a supermassive black hole is currently at the theoretical and simulation levels, and specific conditions need to be satisfied~\cite{Ullio:2001fb}.

\bigskip

\section*{Acknowledgment}
We acknowledge the kind support of the National Science and Technology Council of Taiwan R.O.C., with grant number NSTC 111-2811-M-007-018-MY3.
P.Y.T. acknowledges support from the Physics Division of the National Center for Theoretical Sciences of Taiwan R.O.C. with grant NSTC 114-2124-M-002-003.
Y.M.Y. is supported in part by Grant No. 113J0073I4 and NSTC Grant No. 111B3002I4. and by the doctoral scholarship from the Ministry of Education of Taiwan R.O.C.
The authors would like to thank Koichi Hamaguchi, Volodymyr Takhistov, and Shao-Ping Li for discussions and comments.

\bigskip
\newpage

\appendix
\section{Calculation of Cross Section}
\label{app:A}
\label{app:annihilation}
\begin{figure}
\centering
\subfloat[]
{\includegraphics[width=0.538\linewidth]{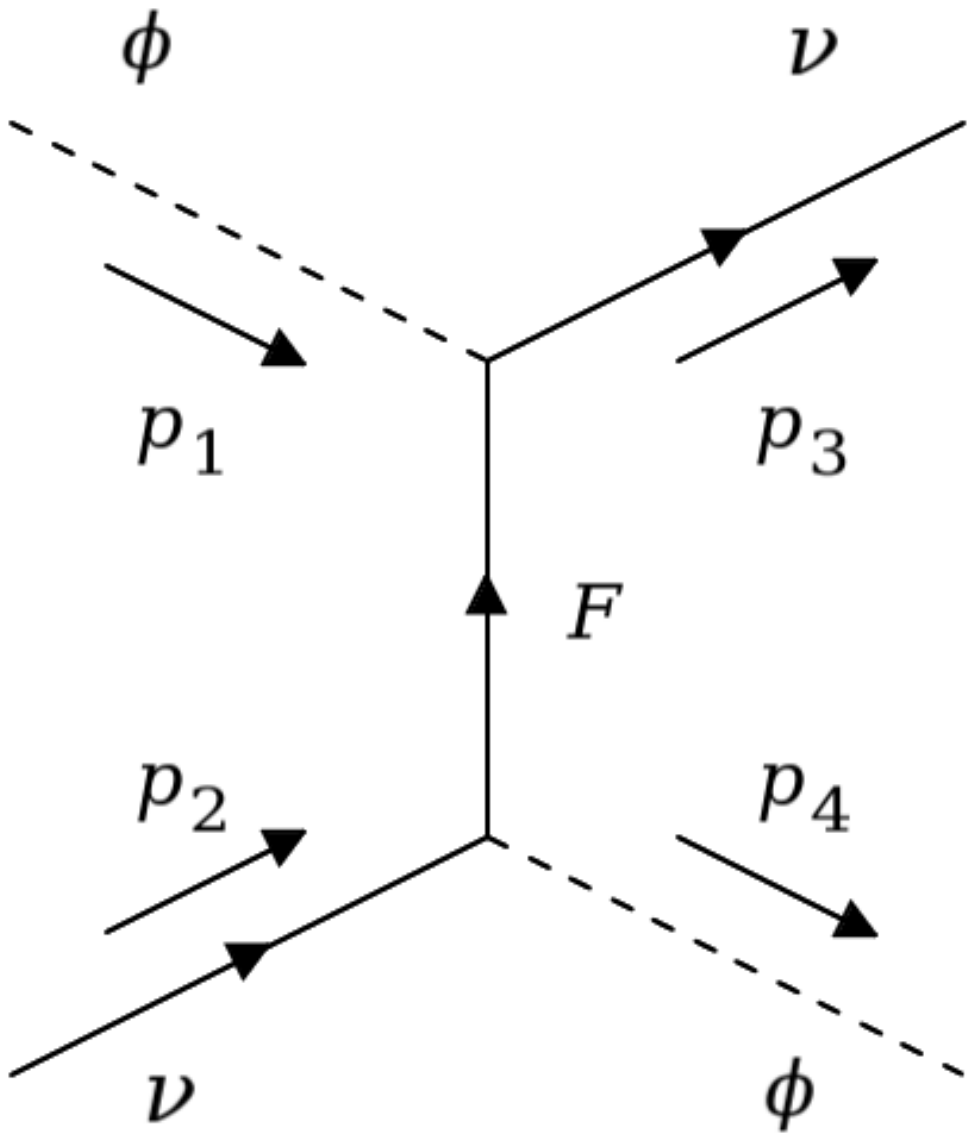}}
\caption{The Feynman diagram of $\nu \phi$ scattering cross section.}
\label{fig:feynman2}
\end{figure}

 For the interaction (\ref{newinteraction}), the amplitude is given by \cite{Boehm:2003hm} 
\begin{equation}\label{amplitudeF}
\sum_{s_2,s_3}|\mathcal{M}|^2=\frac{4y^4}{(t-m_F)^2}\left[(p_1\cdot p_2)(p_1\cdot p_3)-\frac{m^2_\phi}{2}(p_2\cdot p_3)\right]=\frac{y^4(m_\phi^4-st)}{(t-m_F^2)^2}
\end{equation}
and the Feynman diagram is shown in Fig.  \ref{fig:feynman2}.
The Mandelstam variable $t$ is given by
	\begin{equation}
	t=\frac{2sm^2_\phi-s^2+m^4_\phi}{2s}-\frac{(s-m^2_\phi)^2}{2s}\cos\theta\equiv A-B\cos\theta,
	\end{equation}
where $\theta$ is the scattering angle in the CM frame. Then (\ref{amplitudeF}) becomes
\begin{equation}
\sum_{s_2,s_3}|\mathcal{M}|^2=\frac{y^4[m^4_\phi-s(A-B\cos\theta)]}{(A-m_F^2-B\cos\theta)^2}.
\end{equation}
The cross section is
\begin{equation}\label{nuphicrosssection2}
\sigma_{\nu\phi}=\frac{I}{64\pi^2 s},
\end{equation}
where
\begin{equation}\label{a.23}
I=\int d\Omega \left\langle \sum_{s_2,s_3}|\mathcal{M}|^2\right\rangle= y^4\pi\left\{
\frac{s[(m_F^2-A)^2-B^2]\ln\left(\left|\frac{m_F^2+B-A}{m_F^2-B-A}\right|\right)+2B(m^4_\phi-sm_F^2)}{B[(m_F^2-A)^2-B^2]}
\right\}.
\end{equation}
The neutrinos from the AGN have energy of TeV scale, which is much greater than the DM mass $m_\phi$, thus in the $\phi$ rest frame
\begin{subequations}
\begin{align}
	s&=m^2_\phi+2E_\nu m_\phi\simeq 2E_\nu m_\phi,\quad A\simeq -E_\nu m_\phi\simeq-B
\end{align}
\end{subequations}
In this case (\ref{a.23}) can be approximated by 
	\begin{equation}
	I\simeq 2y^4\pi\left[\ln\left(1+\frac{2E_\nu m_\phi}{m^2_F}\right)-\frac{2E_\nu m_\phi }{m^2_F+2E_\nu m_\phi }\right]
	\end{equation}
and the corresponding cross section is
\begin{equation}\label{nuphicrosssection3}
	\sigma_{\nu\phi}=\frac{y^4}{32\pi}\left[\frac{\ln\left(1+\frac{2E_\nu m_\phi}{m^2_F}\right)}{2E_\nu m_\phi}-\frac{1}{m^2_F+2E_\nu m_\phi }\right].
\end{equation}
In our analysis, we set $m_F\geq 1\,{\rm TeV}$ and $m_F^2\gg E_\nu m_\phi$. The cross section then becomes linear in $E_\nu$:
\begin{equation}
\sigma_{\nu\phi}=\left(\frac{y^4m_\phi}{32\pi m^4_F}\right)E_\nu.
\end{equation}

As For the $\phi\phi^\ast$ annihilation, we consider the nonself conjugate scalar DM annihilation,
the corresponding amplitude is given by \cite{Boehm:2003hm}
	\begin{align}
	\int^1_{-1}|\mathcal{M}|^2d\cos\theta=\frac{16y^4|{\bf p}_i|^2 m_\phi^2}{3(m_F^2+m^2_\phi)^2}.
	\end{align} 
So the total cross section is 
	\begin{equation}
	\sigma_{\phi\phi}=\frac{1}{32\pi s}\frac{|{\bf p}_f|}{|{\bf p}_i|}\int^1_{-1}|\mathcal{M}|^2d\cos\theta.
	\end{equation}
The amplitude of initial and final three momentum $|{\bf p}_i|$ is
	\begin{equation}
	|{\bf p}_i|=m_\phi\gamma \langle v\rangle,\quad |{\bf p}_f|=\sqrt{m^2_\phi\gamma^2-m_\nu^2}
	\end{equation}
with $\gamma=(1-\langle v\rangle^2)^{-1/2}$, $s=(2m_\phi\gamma)^2$.

The differential cross section in the lab frame is given by Eq.(9) of Ref.~\cite{Arguelles:2017atb}
\begin{equation}\label{diffxsec}
    \frac{d\sigma}{dx}=\frac{1}{32\pi m_\phi E_\nu}\frac{E'^2_\nu}{E_\nu m_\phi}\left\langle \sum_{s_2,s_3}|\mathcal{M}|^2\right\rangle,
\end{equation}
where $x=\cos\theta$, $E_\nu$ is the incident neutrino energy, $E'_\nu$ is the scattered energy, and they are related by
\begin{equation}
\frac{1}{E'_\nu}=\frac{1}{E_\nu}+\frac{1-x}{m_\phi}.
\end{equation}
We can change the differential variable of (\ref{diffxsec}) by
    \begin{equation}
    \frac{d\sigma}{dE'_\nu}=\frac{1}{32\pi m_\phi E^2_\nu}\left\langle \sum_{s_2,s_3}|\mathcal{M}|^2\right\rangle.
    \end{equation}

\bigskip

\section{The Detail Calculation of $\nu{\rm Ar}$ Scattering}
\label{appendix:B}
The cross section of the charged current $\nu_e\prescript{40}{}{{\rm Ar}}$ scattering is given by 
\begin{align}
	\sigma_{\nu {\rm Ar}}&=\frac{G^2_F|V_{ud}|^2 E^{\rm CM}_e|{\bf p}_e^{\rm CM}|}{\pi}\left[\frac{(\sqrt{s}-E_e^{\rm CM})E^{\rm CM}_{\rm Ar}}{s}\right]F_C\left[B({\rm F})+B({\rm GT})\right],
\end{align}
where $G_F=1.17\times 10^{-5}\,{\rm GeV}^{-2}$ 
is the Fermi constant, $V_{ud}$ is the CKM matrix element connecting the up and down quarks. $F_C$ is the allowed approximation Coulomb correction factor \cite{Gardiner:2018zfg}
	\begin{equation}
	F_C=\begin{cases}
		F(Z_f,v_{\rm rel}),\  f^2_{\rm EMA}>F(Z_f,v_{\rm rel})\\
		f^2_{\rm EMA},\ \text{otherwise}
	\end{cases},
	\end{equation}
where
	\begin{equation}\label{b.3}
		F(Z_f,E_e^{\rm FNR})=\frac{2(1+S)}{[\Gamma(1+2S)]^2}(2|{\bf p}^{\rm FNR}_e|R)^{2S-2}e^{-\pi\eta}|\Gamma(S+i\eta)|^2
	\end{equation}
is the Fermi function in the "final nucleus rest frame" (FNR frame). 
We transform the four momenta in the CM frame into the FNR frame. 
In the CM frame, the four momentum of $e^-$ and $^{40}{\rm K}$ are $p_e=(E^{\rm CM}_e,{\bf p}^{\rm CM}_e)$ and $p_{\rm K}=(E^{\rm CM}_{\rm K},-{\bf p}^{\rm CM}_e)$
with
	\begin{equation}\label{b.6}
	E^{\rm CM}_e=\frac{s+m^2_e-m^2_{\rm K}}{2\sqrt{s}},\quad E^{\rm CM}_{\rm K}=\frac{s-m^2_e+m^2_{\rm K}}{2\sqrt{s}}.
\end{equation}	
In the FNR frame, $^{40}{\rm K}$ is at rest, so
	\begin{equation}
	{\rm p}^{\rm FNR}_{\rm K}\hat{\bf x}=\gamma(-|{\bf p}^{\rm CM}_e|-E^{\rm CM}_{\rm K}v)\hat{\bf x}={\bf 0},
\end{equation}
where $\hat{\bf x}$ is the direction of electron.
This yields $v=-{|{\bf p}^{\rm CM}_e|}/{E^{\rm CM}_{\rm K}}$. 
The four momentum of electron in FNR frame is 
	\begin{equation}
	\begin{aligned}
		{\rm p}^{\rm FNR}_e\hat{\bf x}&=\gamma\left(|{\bf p}^{\rm CM}_e|+E_e^{\rm CM}\frac{|{\bf p}^{\rm CM}_e|}{E^{\rm CM}_{\rm K}}\right)\hat{\bf x}
		=\gamma|{\bf p}^{\rm CM}_e|\left(1+\frac{E_e^{\rm CM}}{E^{\rm CM}_{\rm K}}\right)\hat{\bf x},\\
		E^{\rm FNR}_e&=\gamma\left(E^{\rm CM}_e+|{\bf p}^{\rm CM}_e|\frac{|{\bf p}^{\rm CM}_e|}{E^{\rm CM}_{\rm K}}\right)
		=\gamma\left(E^{\rm CM}_e+\frac{|{\bf p}^{\rm CM}_e|^2}{E^{\rm CM}_{\rm K}}\right).
	\end{aligned}
	\end{equation}
The relative velocity of electron to the rest ${\rm K}$ is then $v_{\rm rel}={{\rm p}^{\rm FNR}_e}/{E^{\rm FNR}_e}$, and the Lorentz factor is $\gamma_{\rm rel}={E^{\rm FNR}_e}/{m_e}$. Hence we may write (\ref{b.3}) as 
		\begin{equation}
		F(Z_f,v_{\rm rel})=\frac{2(1+S)}{[\Gamma(1+2S)]^2}(2\gamma_{\rm rel}v_{\rm rel}m_e R)^{2S-2}e^{-\pi\eta}|\Gamma(S+i\eta)|^2.
	\end{equation}
The velocity of the CM frame observed from the lab (${\rm Ar}$ rest) frame is given by
\begin{equation}
	v_{\rm CM}=\frac{{\rm p}_{\nu x}}{E_\nu+m_{\rm Ar}}=\frac{|{\bf p}_\nu|}{E_\nu+m_{\rm Ar}}
\end{equation}
and
\begin{equation}
	\gamma_{\rm CM}=\frac{1}{\sqrt{1-v_{\rm CM}^2}}=\frac{m_{\rm Ar}+E_\nu}{\sqrt{m^2_{\rm Ar}+2m_{\rm Ar}E_\nu}}.
\end{equation}
The Mandelstam variable $s$ in the lab frame is
	\begin{equation}\label{b.18}
	s=m_{\rm Ar}^2+2E_\nu m_{\rm Ar}.
	\end{equation}
With (\ref{b.18}) and (\ref{b.6}) we can write $E^{\rm CM}_e$, $|{\bf p}^{\rm CM}_e|$ in terms of masses and $E_\nu$, and $E^{\rm CM}_{\rm Ar}=\gamma_{\rm CM}m_{\rm Ar}$.

$f_{\rm EMA}$ is the rescaled factor of effective momentum approximation (EMA) and is given by 
	\begin{equation}
	f_{\rm EMA}=\frac{|{\bf p}^{\rm eff}_e|}{|{\bf p}_e|},
	\end{equation}
where
	\begin{equation}
	|{\bf p}^{\rm eff}_e|=\sqrt{\left(E_e+\frac{3Z_f\alpha}{2R}\right)^2-m^2_e}.
	\end{equation}

\bigskip

\section{Calculation for $\rho_0$ and $r_0$}
\label{appendix:C}

We follow the formula in \cite{Gentile:2007sb}, the characteristic radius and density of the distribution are given by
	\begin{align}\label{c.1}
	\rho_0=\frac{\Delta}{3}\frac{c^3}{\ln(1+c)-\frac{c}{1+c}}\rho_c,\quad
	r_0\simeq 8.8\left(\frac{M_{\rm vir}}{10^{11}M_\odot}\right)^{0.46}\,{\rm kpc}
	\end{align}
where $\Delta=200$ is the virial overdensity and 
	\begin{subequations}\label{c.3}
	\begin{align}
		&\rho_c=1.053672\times10^{-5}\, h^2\,({\rm GeV}/c^2){\rm cm}^{-3}\simeq 4.78658\times10^{-6}\,({\rm GeV}/c^2){\rm cm}^{-3},\\
		&c\simeq 13.6\left(\frac{M_{\rm vir}}{10^{11}M_\odot}\right)^{-0.13}
	\end{align}
	\end{subequations}
are the critical density of the universe and concentration parameter.
The DM halo mass is related to the central supermassive black hole mass by \cite{Cline:2023tkp}
	\begin{equation}
	M_{\rm DM}\sim 10^{12}M_\odot\times\left(\frac{M_{\rm BH}}{7\times10^7 M_\odot}\right)^{3/4}.
	\end{equation}
We take the DM halo mass to be the virial mass, then we have
	\begin{equation}
	\begin{aligned}
	M_{\rm DM}\simeq 2.32\times 10^{11} M_\odot \quad({\rm NGC}),\\
	M_{\rm DM}\simeq 3.05\times 10^{12} M_\odot \quad({\rm TXS}),\\
	\end{aligned}
	\end{equation}
and from (\ref{c.1})-(\ref{c.3}) we have
	\begin{equation}
	\begin{aligned}
	&\rho_0\simeq0.35\,{\rm GeV}/{\rm cm}^3,\quad r_0\simeq 13\,{\rm kpc}\quad({\rm NGC}),\\
	&\rho_0\simeq0.154\,{\rm GeV}/{\rm cm}^3,\quad r_0\simeq 42.36\,{\rm kpc}\quad({\rm TXS}).
	\end{aligned}
	\end{equation}

\bigskip
 

\end{document}